\documentclass[conference]{IEEEtran}

\usepackage{cite}
\usepackage{amsmath,amssymb,amsfonts}
\usepackage{algorithmic}
\usepackage{graphicx}
\usepackage{textcomp}
\usepackage{xcolor}
\usepackage{tikz}
\usepackage{float}
\usepackage{comment}
\usepackage{paralist}
\usepackage{enumitem}
\usepackage{xspace}
\usepackage{svg}
\usepackage{algorithm}
\usepackage{algorithmic}
\usepackage{adjustbox}
\usepackage{booktabs}
\usepackage{url}
\usepackage{courier}
\usepackage{circledsteps}
\usepackage[numbers,compress]{natbib}
\usepackage[colorlinks, linkcolor=black, citecolor=black, urlcolor=blue]{hyperref}
\usetikzlibrary{shadows}
\usepackage{numprint}
\npdecimalsign{.}
\nprounddigits{0}
\usepackage{pgf}

\def\BibTeX{{\rm B\kern-.05em{\sc i\kern-.025em b}\kern-.08em
    T\kern-.1667em\lower.7ex\hbox{E}\kern-.125emX}}

\newcommand{\vizName}{\textsc{EvoScat}\xspace}
\newcommand{\toolURL}{\url{https://design.inf.usi.ch/evoscat}}

\newcommand{\legendbox}[1]{
\resizebox{!}{1em}{%
\begin{tikzpicture}
\definecolor{myyellow}{HTML}{#1}
    \draw[fill=myyellow, draw=black] (0,0) rectangle (0.5,0.5);
\end{tikzpicture}}
}

\makeatletter
\newcommand{\linebreakand}{%
  \end{@IEEEauthorhalign}
  \hfill\mbox{}\par
  \mbox{}\hfill\begin{@IEEEauthorhalign}
}
\makeatother

\pgfkeys{/csteps/inner ysep=5pt}
\pgfkeys{/csteps/inner xsep=5pt}
\pgfkeys{/csteps/inner color=white}
\pgfkeys{/csteps/fill color=black}
\pgfkeys{/csteps/outer color=black}

\usepackage{scalerel}
\usetikzlibrary{svg.path}

\definecolor{orcidlogocol}{HTML}{A6CE39}
\tikzset{
  orcidlogo/.pic={
    \fill[orcidlogocol] svg{M256,128c0,70.7-57.3,128-128,128C57.3,256,0,198.7,0,128C0,57.3,57.3,0,128,0C198.7,0,256,57.3,256,128z};
    \fill[white] svg{M86.3,186.2H70.9V79.1h15.4v48.4V186.2z}
                 svg{M108.9,79.1h41.6c39.6,0,57,28.3,57,53.6c0,27.5-21.5,53.6-56.8,53.6h-41.8V79.1z M124.3,172.4h24.5c34.9,0,42.9-26.5,42.9-39.7c0-21.5-13.7-39.7-43.7-39.7h-23.7V172.4z}
                 svg{M88.7,56.8c0,5.5-4.5,10.1-10.1,10.1c-5.6,0-10.1-4.6-10.1-10.1c0-5.6,4.5-10.1,10.1-10.1C84.2,46.7,88.7,51.3,88.7,56.8z};
  }
}

\newcommand\orcidicon[1]{\href{https://orcid.org/#1}{\mbox{\scalerel*{
\begin{tikzpicture}[yscale=-1,transform shape]
\pic{orcidlogo};
\end{tikzpicture}
}{|}}}}

\begin{document}

\title{\vizName: Exploring Software Change Dynamics in Large-Scale Historical Datasets}

\author{
    \IEEEauthorblockN{Souhaila Serbout~\orcidicon{0000-0002-8144-2606}}
    \IEEEauthorblockA{\emph{University of Zurich} \\
    Zurich, Switzerland \\
    souhaila.serbout@uzh.ch}
    \and
    \IEEEauthorblockN{Diana Carolina Muñoz Hurtado~\orcidicon{0000-0002-4769-3444}}
    \IEEEauthorblockA{\emph{Software Institute} \\
    \emph{USI}
    Lugano, Switzerland \\
    carolina.munoz@usi.ch}
    \and
    \IEEEauthorblockN{Hassan Atwi~\orcidicon{0009-0009-7989-5030}}
    \IEEEauthorblockA{\emph{Software Institute} \\
    \emph{USI}
    Lugano, Switzerland \\
    hassan.atwi@usi.ch}
    \and

    \linebreakand

    \IEEEauthorblockN{Edoardo Riggio~\orcidicon{0009-0000-5649-3412}}
    \IEEEauthorblockA{\emph{Software Institute} \\
    \emph{USI}
    Lugano, Switzerland \\
    edoardo.riggio@usi.ch}
    \and
    \IEEEauthorblockN{Cesare Pautasso~\orcidicon{0000-0002-2748-9665}}
    \IEEEauthorblockA{\emph{Software Institute} \\
    \emph{USI}
    Lugano, Switzerland \\
    c.pautasso@ieee.org}
}

\maketitle

\begin{abstract}
Long lived software projects encompass a large number of artifacts, which undergo many revisions throughout their history.
Empirical software engineering researchers studying software evolution gather and collect datasets with millions of events, representing changes introduced to specific artifacts.
In this paper, we propose \vizName, a tool that attempts addressing temporal scalability through the usage of interactive density scatterplot to provide a global overview of large historical datasets mined from open source repositories in a single visualization. \vizName intents to provide researchers with a mean to produce scalable visualizations that can help them explore and characterize evolutions datasets, as well as comparing the histories of individual artifacts, both in terms of 1) observing how rapidly different artifacts age over multiple-year-long time spans 2) how often metrics associated with each artifacts tend towards an improvement or worsening.
The paper shows how the tool can be tailored to specific analysis needs (pace of change comparison, clone detection, freshness assessment) thanks to its support for flexible configuration of history scaling and alignment along the time axis, artifacts sorting and interactive color mapping, enabling the analysis of millions of events obtained by mining the histories of tens of thousands of software artifacts. We include in this paper a gallery showcasing datasets gathering specific artifacts (OpenAPI descriptions, GitHub workflow definitions) across multiple repositories, as well as diving into the history of specific popular open source projects.
\end{abstract}


\section{Introduction}

As software systems evolve, they leave behind a rich trail of events that captures their transformation over time. Historical datasets include changes to the software as well as its accompanying test cases, documentation, configuration files, and their version tracking metadata~\cite{fischer2003ICSM}. Researchers mine this data to analyze software evolution patterns~\cite{novais2013software,bohner2007evolutional}, uncover trends~\cite{sousa2022time}, and develop tools for tracking and improving software quality and maintainability~\cite{broneske2024sharing,hecht2015tracking,hurtado2025mining,fischer2006evograph,ogawa2008stargate, ambros2005fractal,telea2005interactive}.

Drawing insights from such rich historical data requires appropriate means of exploration, as the complexity and volume of versioned software artifacts can quickly become overwhelming. Among these, visualization plays a crucial role~\cite{pinzger2005visualizing,wu2004evolution,ambros2005fractal,telea2005interactive}. Visualizing evolving parts of software projects at scale helps to compare various artifacts within the same view, which can facilitate the detection of similarities, divergence, and time related trends.  However, a critical challenge emerges when attempting to scale these visualizations to datasets containing evolution data of a large number of projects or to projects that produce millions of data points tracking each component's evolutionary steps~\cite{richer2022scalability}.
Scatter plots can efficiently represent a large number of data points within the same view, allowing visual comparison based on data points color, positions, shapes and density~\cite{hilasaca2023grid,mayorga2013splatterplots,wang2017line}. They allow multiple dimensions of information to be encoded simultaneously: the position of each dot can represent attributes such as time and artifact identity, and the color can encode change types, their age, or artifact categories~\cite{wang2017line,hilasaca2023grid}.

In the context of software evolution, we designed \vizName, a large scale data visualization tool that leverages scatter plots exploiting their visual encoding flexibility. \vizName uses this flexibility by mapping time to one of the axis, allowing the viewer to follow the temporal evolution of software components vertically. Each artifact history is represented as a vertical sequence of dots, where each dot corresponds to a specific event (e.g., commit) in the artifact's life time. The other dimension is used to organize artifacts---by name, directory structure, similarity, or logical grouping---so that patterns such as synchronized changes~\cite{bala2017uncovering}, divergence, or inactivity across components become visually salient. The key challenge of using a scatter plot to visually represent and compare the histories of thousands of artifacts concerns how these are sorted to be placed along the artifact axis.

In this paper, we showcase how \vizName can be used to create such a visualization in the case of eight real-world large-scale software evolution datasets containing different types of software artifacts---code files, CI/CD pipeline configurations file, API documentation files---and covering two levels of granularity---across artifacts of the same kind mined from multiple open-source repositories and across all artifacts found within the same repository.

This paper makes the following contributions: 1) we discuss the design space of the evolution scatterplot addressing the use cases of empirical software engineering researchers; 2) we introduce the interactive visualization tool\footnote{\toolURL} and the filtering features it allows it supports to generate visualizations tailored to the analysis needs; 3) we showcase a gallery of diverse historical datasets visualizations generated by \vizName, attempting to reveal some insights about key events occurring during the evolution of these software projects.

The rest of this paper is organized as follows. In Section~\ref{sec:req}
we collect the use cases which drive the design of \vizName's visualization. In Section~\ref{sec:design} we present tool's filtering features and its implementation details. We showcase the usage of \vizName on a selection of historical datasets mined from open source repositories in Section~\ref{sec:gallery} and present a reflection on the scalability and limitations of \vizName in Section~\ref{sec:discuss}. Before drawing some conclusions in Section~\ref{sec:theend}, we introduce the related work in Section~\ref{sec:relw}.

\section{Use Cases}\label{sec:req}

To guide the design of the visualization tool, we identified a set of key use cases that reflect the challenges faced when analyzing large-scale software evolution datasets~\cite{richer2022scalability,novais2013software}. These use cases are derived from practical analysis tasks and recurring needs in empirical software engineering research~\cite{bani2016software,lowe2020requirements}, and they ensure \vizName's temporal scalability for evolution data belonging different artifact types and project contexts.
\vspace{.2cm}

\noindent \textbf{Dataset Overview \& Temporal Coverage}
\begin{enumerate}[label=U\arabic*., start=1]
    \item Provide an overview of large historical datasets supporting its initial exploration (Figs.~\ref{plot:ghwf_commit_absolute_time},~\ref{plot:plot_comparison}).
    \item Assess the uniform or skewed distribution of event occurrence over time and the artifacts' age across the sampling time period of the dataset (Figs.~\ref{plot:oas5-first-absolute},~\ref{plot:ghwf_commit_absolute_time},~\ref{plot:oas_complete_filter}).

\end{enumerate}

\noindent{\textbf{Artifact Evolution \& Development Trends Comparison}}
\begin{enumerate}[label=U\arabic*., start=3]
    \item Compare the development pace or rhythm of different artifacts belonging to the same or different projects (Fig.~\ref{plot:vscode_id_absolute_time}).
    \item Distinguish artifacts undergoing high intensity effort vs. regular maintenance (Fig.~\ref{plot:vscode_id_absolute_time}).
    \item Highlight the stability or variability of metrics associated with each commit, separating the artifacts where the metric improves over time from the ones with degraded measurements (Fig.~\ref{plot:ghwf_misconfiguration_delta_first_last}).
\end{enumerate}

\noindent\textbf{Anomaly \& Pattern Discovery}
\begin{enumerate}[label=U\arabic*., start=6]
    \item Help to visually spot forks, clones or duplicated artifacts in the dataset (Fig.~\ref{plot:oas5-first-absolute}).
    \item Hunt for clusters of interesting, outstanding or peculiar artifact evolution histories, to be further analyzed (Figs.~\ref{plot:vscode_id_absolute_time},~\ref{plot:cpython_last_size}).
\end{enumerate}

\begin{figure}[t]
    \centering
    \href{https://design.inf.usi.ch/evoscat/?dataset=vscode#time=absolute&artifact=methods_count_last&color=metric}{\includegraphics[width=.49\textwidth]{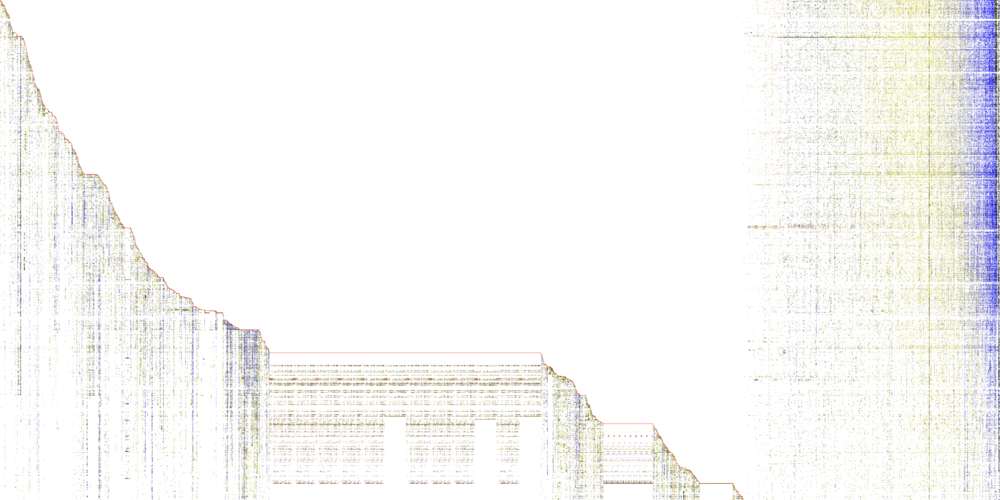}}

    \caption{Evolution Plot of the VS Code Dataset (Artifacts sorted by 1) their size measured in the last commit 2) last commit timestamp, Time is absolute on the vertical axis, Colored by size - \protect\legendbox{FF0000} means the file has been deleted) -- The discarded artifacts throughout the project history pile up on the left, while artifacts that are still present in the repository are placed from small to large on the right side.}
    \label{plot:vscode_last_size}
\end{figure}

\noindent\textbf{Lifecycle \& Activity Monitoring} % Dynamics Monitoring ?
\begin{enumerate}[label=U\arabic*., start=8]
    \item Check the freshness of the artifacts (Figs.~\ref{plot:vscode_last_size},~\ref{plot:evoscat-postgres-double2_delta2-absolute}): has their development been discontinued many years ago? Are there recent changes? What is the ratio between discarded artifacts, stable artifacts, and recently added ones?
\end{enumerate}

\noindent\textbf{Data Volume Scalability}
\begin{enumerate}[label=U\arabic*., start=9]
    \item Scale to visualize millions of events collected, tracking the history of thousands of artifacts over decades (Figs.~\ref{plot:oas5-first-absolute},~\ref{plot:ghwf_commit_absolute_time},~\ref{plot:plot_comparison}–\ref{plot:evoscat-postgres-mid-absolute-delta}). The visualization is intended to represent large datasets using high resolution displays. It is not designed for small sets of short lived artifacts.
\end{enumerate}

\noindent\textbf{Crawler Evaluation \& Dataset Discovery}
\begin{enumerate}[label=U\arabic*., start=10]
    \item Compare the performance of different crawlers regarding their ability to discover previously unknown artifacts vs. how closely they track the evolution of known ones (Figs.~\ref{plot:oas5-first-absolute},~\ref{plot:ghwf_commit_absolute_time}).
\end{enumerate}

\section{Visualization Design}\label{sec:design}

\definecolor{cy2014}{HTML}{FF4A46}
\definecolor{cy2015}{HTML}{FF34FF}
\definecolor{cy2016}{HTML}{FFFF00}
\definecolor{cy2017}{HTML}{008941}
\definecolor{cy2018}{HTML}{1966FF}
\definecolor{cy2019}{HTML}{1CFFD9}
\definecolor{cy2020}{HTML}{C00069}
\definecolor{cy2021}{HTML}{FFDBE5}
\definecolor{cy2022}{HTML}{FF9900}
\definecolor{cy2023}{HTML}{8148D5}
\definecolor{cy2024}{HTML}{FF0066}

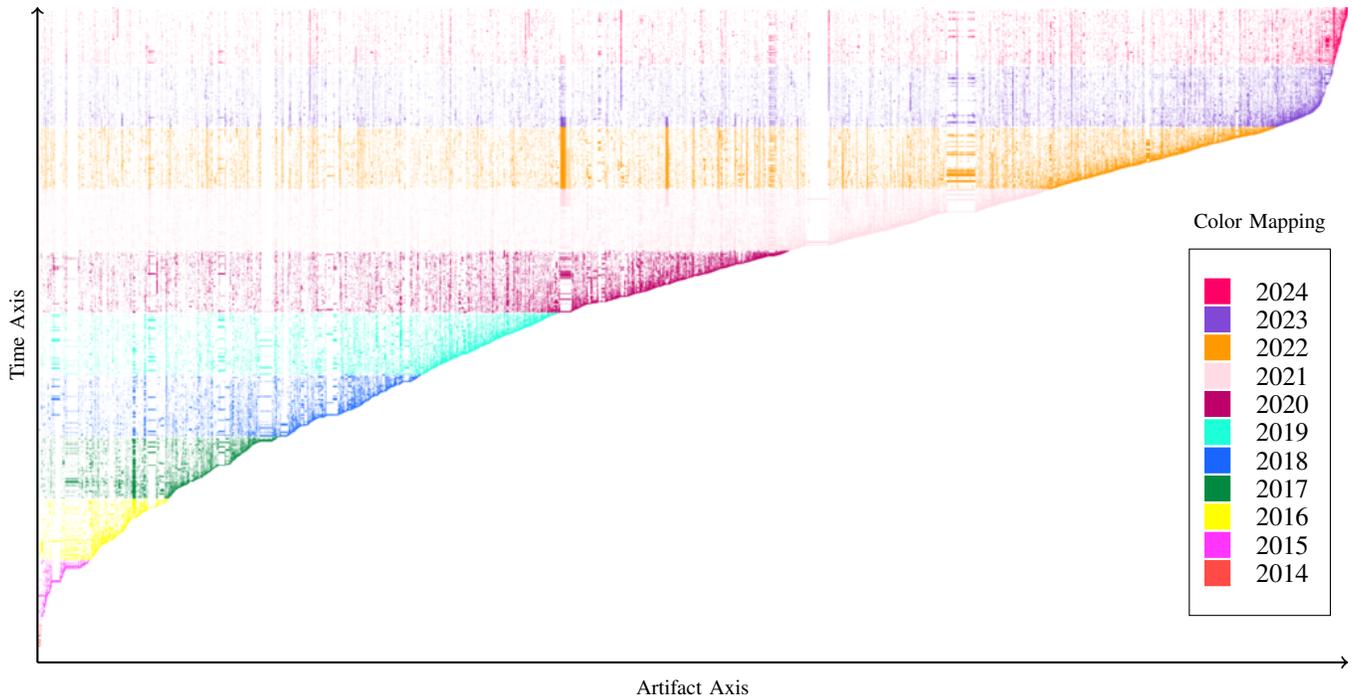
\begin{figure*}[!t]

\begin{tikzpicture}
  % Include the image
  \node[anchor=south west, inner sep=0] (image) at (0,0) {\includegraphics[width=0.96\textwidth]{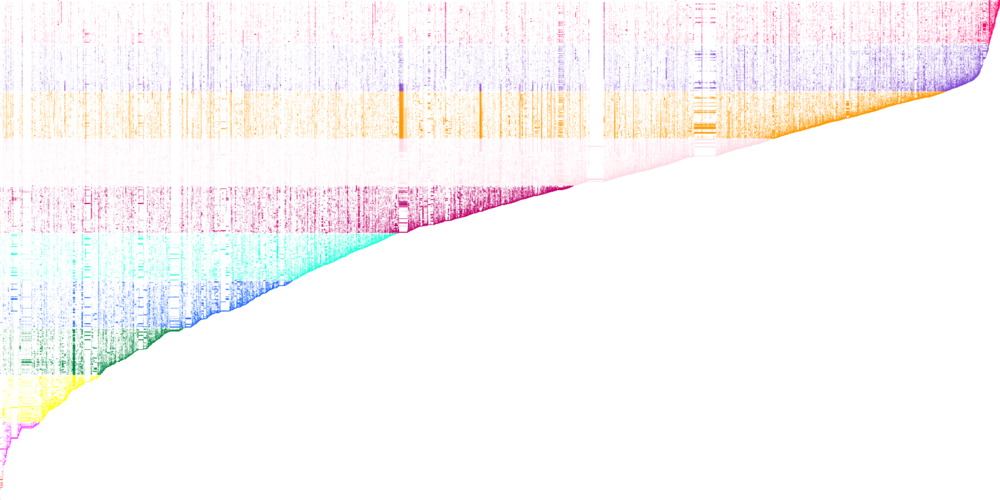}};

  % Get image dimensions
  \begin{scope}
    % Draw X axis
    \draw[->, thick] (image.south west) -- (image.south east)
      node[midway, below=1mm] {\footnotesize Artifact Axis};

    % Draw Y axis
    \draw[->, thick] (image.south west) -- (image.north west)
      node[midway, rotate=90, below=-0.5cm] {\footnotesize Time Axis};
  \end{scope}

  \begin{scope}[shift={(3.5cm,1cm)}, scale=0.75]

    \draw[draw=black] (15.75,6) rectangle (18.25,-0.5);
    \node[above=0.1cm] at (17,6) {\footnotesize Color Mapping};

    \draw[fill={cy2024}, draw=white, thick] (16,5) rectangle (16.5,5.5);
    \node[right=0.2cm] at (16.5,5.25) {2024};
    \draw[fill={cy2023}, draw=white, thick] (16,4.5) rectangle (16.5,5);
    \node[right=0.2cm] at (16.5,4.75) {2023};
    \draw[fill={cy2022}, draw=white, thick] (16,4) rectangle (16.5,4.5);
    \node[right=0.2cm] at (16.5,4.25) {2022};
    \draw[fill={cy2021}, draw=white, thick] (16,3.5) rectangle (16.5,4);
    \node[right=0.2cm] at (16.5,3.75) {2021};
    \draw[fill={cy2020}, draw=white, thick] (16,3) rectangle (16.5,3.5);
    \node[right=0.2cm] at (16.5,3.25) {2020};
    \draw[fill={cy2019}, draw=white, thick] (16,2.5) rectangle (16.5,3);
    \node[right=0.2cm] at (16.5,2.75) {2019};
    \draw[fill={cy2018}, draw=white, thick] (16,2) rectangle (16.5,2.5);
    \node[right=0.2cm] at (16.5,2.25) {2018};
    \draw[fill={cy2017}, draw=white, thick] (16,1.5) rectangle (16.5,2);
    \node[right=0.2cm] at (16.5,1.75) {2017};
    \draw[fill={cy2016}, draw=white, thick] (16,1) rectangle (16.5,1.5);
    \node[right=0.2cm] at (16.5,1.25) {2016};
    \draw[fill={cy2015}, draw=white, thick] (16,0.5) rectangle (16.5,1);
    \node[right=0.2cm] at (16.5,0.75) {2015};
    \draw[fill={cy2014}, draw=white, thick] (16,0) rectangle (16.5,0.5);
    \node[right=0.2cm] at (16.5,0.25) {2014};
  \end{scope}

\end{tikzpicture}
    \caption{Evolution Plot of the Filtered OpenAPI Dataset (Artifacts sorted by first commit timestamp, Absolute time on the vertical axis, Colored by commit year). This provides insight into the adoption of the OpenAPI standard, with early adopters positioned on the bottom left corner and highlighting the yearly increase in usage, as reflected from the OpenAPI specifications found on GitHub open source repositories.}
    \label{plot:oas5-first-absolute}
\end{figure*}

\vizName uses a simple density scatter plot, where each dot represents an event in the history of an artifact evolution, as tracked by a version control repository. The events belonging to the history of the same artifacts are positioned on the same vertical line. The initial event at the bottom of the dotted line represents the creation of the artifact, which may be changed by committing new revisions to the repository at a specific timestamp, until it is removed (or no more changes occur).
In this section, we define the various configuration options available in the current version of \vizName, enabling users to tailor the visualization to their specific analysis needs.

The vertical axis of the visualization represents time, while the other indicates the specific artifact being tracked (Fig.~\ref{plot:oas5-first-absolute}). We choose the Y axis as the time axis in order to visually represent the stratification of time from bottom to top, so that newer artifact versions layer on top of previous ones. This also leaves the X axis to somehow place the artifacts side-by-side so that their interdependent evolution histories can be tracked by visually following parallel vertical lines.

Given that the plot is meant to represent a very high number of artifacts and events in their history, we only use one dot for each event. Its color can be mapped to summarize relevant properties of the artifact or the changes applied to it in the particular commit. For instance, in the case of Figs.~\ref{plot:oas5-first-absolute} and~\ref{plot:ghwf_commit_absolute_time}, the color is used to distinguish the commits by their corresponding years. This helps to get an overview of the widespread usage of GitHub workflow in public software repositories over the years, since the technology was introduced.

While it would be possible to also use the dot shape or size to visually map further information from the dataset, such details would not be clearly visible in the overview when the entire dataset of a large number of events is displayed.
We also do not draw explicitly the axes because, unlike traditional scatter plots emphasizing the relationships between the independent variable (X) and the one depending on it (Y), whose values are read on linear or log scales, in our case we focus on the resulting visualization shape, which should visually represent the similarities (or differences) between the artifact evolution histories. It would also not be feasible to print the identifiers of a very large number of artifacts positioned side by side along the X axis. To reify the visualization, we show the timestamp and the artifact identifier as the user interacts with its elements.

\subsection{Filtering} To ensure a sharp visualization, a filtering mechanism is applied. During pre-processing, it is possible to specify \textit{the minimum number of events per history} to hide the artifacts that lack a sufficient number of data points. The filtering helps to avoid a sparse, low density plot by visualizing only the artifacts with substantial evolution histories.

\subsection{Time Axis Configuration}
\vizName supports various time scales for the Y axis.  We distinguish five possible configurations:

\begin{asparaitem}
\item \textbf{Absolute:} Represents time in its natural form, preserving the actual timestamps of events. This mode is useful for examining changes in real chronological order as time flows from bottom to top, and the most recent events are layered on top of the older ones.

\item \textbf{Relative to the Start:} Sets the initial point of all artifact histories to a common origin. Each artifact timeline is aligned and can be directly compared as it advances with the same speed on each vertical track (Fig.~\ref{plot:ghwf_commit_absolute_time} (d)).

\item \textbf{Relative to the End:} Aligns the end points of artifact histories, facilitating the visual comparison of the most recent changes across multiple artifacts.

\item \textbf{Relative to the Median Date:} Centers the timeline around the median commit date, revealing symmetries or asymmetries in the artifact evolution histories, highlighting collective changes or synchronized evolution phases.

\item \textbf{Normalized:} Scales each timestamp between 0 and 1, normalizing the time domain within which artifacts evolve, stretching it across the entire visualization regardless of their individual lifespan duration. Using this mode, one can see if, across their entire history -- independently of its duration or how many commits are present -- the artifacts have a similar tendency to change (e.g., grow or shrink, or to become more or less error-prone as shown in Fig.~\ref{plot:ghwf_misconfiguration_delta_first_last}).
\end{asparaitem}

\subsection{Artifact Axis Sorting Criteria}

\begin{figure*}[t]
    \centering
    \begin{minipage}{0.49\textwidth}
        \raggedright\textbf{(a)}\\[-3em]
        \href{https://design.inf.usi.ch/evoscat/?dataset=wfgh#time=absolute&artifact=last&color=year}{\includegraphics[width=\textwidth]{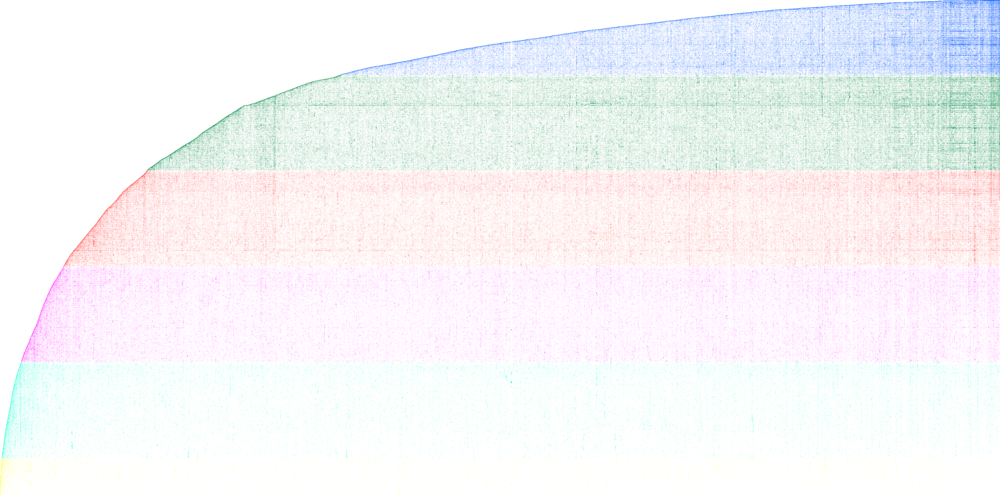}}
    \end{minipage}
    \hspace{0.3mm}
    \begin{minipage}{0.49\textwidth}
        \raggedleft\textbf{(b)}\\[-3em]
        \href{https://design.inf.usi.ch/evoscat/?dataset=wfgh#time=absolute&artifact=first&color=year}{\includegraphics[width=\textwidth]{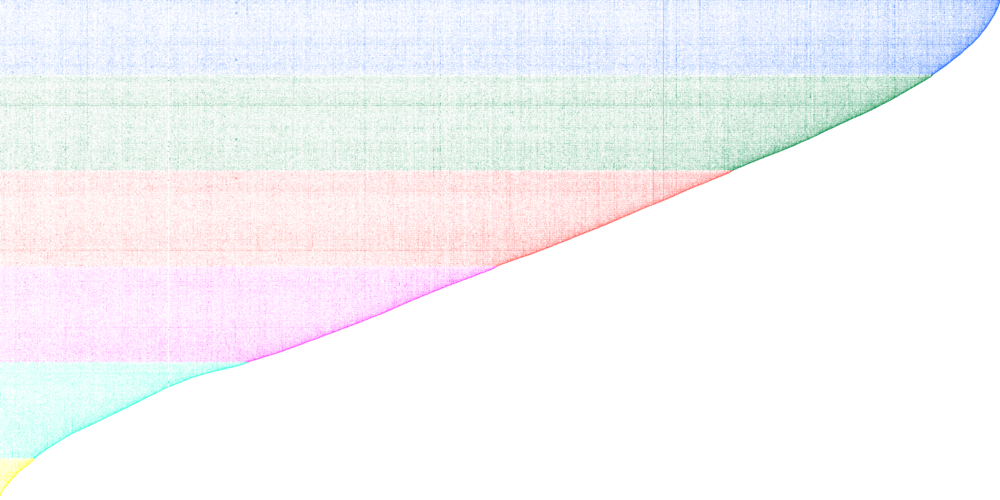}}
    \end{minipage}

    \vspace{2.9em}

    \begin{minipage}{0.49\textwidth}
        \raggedright\textbf{(c)}\\[-3em]
        \href{https://design.inf.usi.ch/evoscat/?dataset=wfgh#time=absolute&artifact=mid&color=year}{\includegraphics[width=\textwidth]{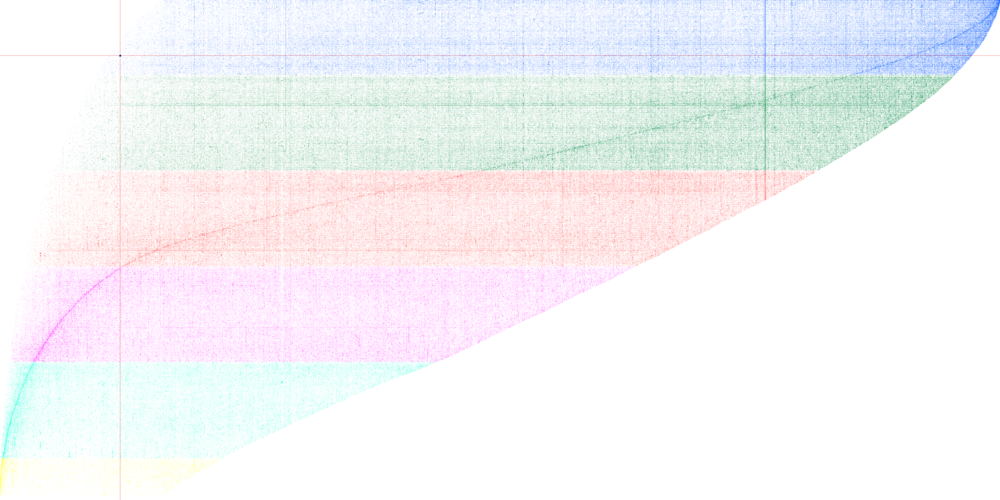}}
    \end{minipage}
    \hspace{0.2mm}
    \begin{minipage}{0.49\textwidth}
       \center\textbf{(d)}\\[-3em]
        \href{https://design.inf.usi.ch/evoscat/?dataset=wfgh#time=normage&artifact=age&color=year}{\includegraphics[width=\textwidth]{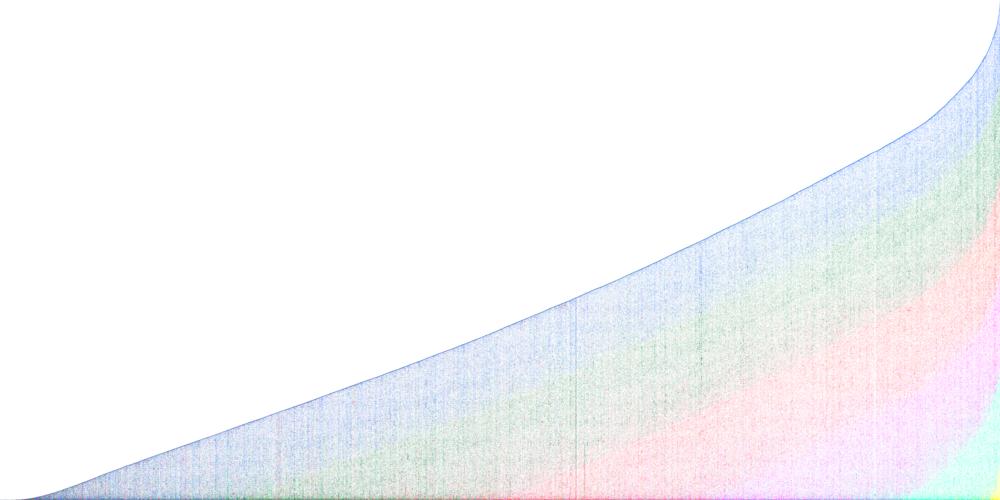}}
    \end{minipage}

    \caption{
        Each evolution scatterplot of the GitHub Workflows Dataset shows the history of artifacts as vertical sequences of commit events (dots), with absolute time on the vertical axis and color encoding the year of each commit: \protect\legendbox{FFFF00} 2019, \protect\legendbox{1CFFD9} 2020,
        \protect\legendbox{FF34FF} 2021,
        \protect\legendbox{FF4A46} 2022,
        \protect\legendbox{008941} 2023,
        \protect\legendbox{1966FF} 2024.
        Artifacts are sorted by their (a) last, (b) first, (c) median commit timestamp, and (d) age (last-first commit).
    }
    \label{plot:ghwf_commit_absolute_time}
\end{figure*}

Artifact histories are displayed as sequences of dots vertically with each artifact side by side in the visualization. The order of the artifacts in the layout is a critical parameter affecting the resulting visualization, as their placement can highlight similarities between artifact histories. Given the high number of artifacts present in the visualization, there is a combinatorial explosion of possible orderings. Also, it is not practical to manually place thousands of artifacts along a certain order.
Therefore, artifacts can be automatically sorted along the X axis according to different criteria, making the visualization very flexible so that it can be easily tailored to specific analysis needs (Fig.~\ref{plot:ghwf_commit_absolute_time}).

\begin{asparaitem}
\item \textbf{By Number of Events:} Groups artifacts by how many events are found in their history. This separates rarely changing artifacts from the ones that underwent intensive modifications during an eventful evolution history.

\item \textbf{By Starting Time:} Orders artifacts by the time of their first recorded event, providing insight into initial development phases. This visualization indicates the rate at which artifacts were created over time, with the oldest on the bottom left and the most recent on the top right. When the curve emerging from the layout flattens, it indicates that at that time, many artifacts have been created (Fig.~\ref{plot:oas5-first-absolute}).

\item \textbf{By Ending Time:} Prioritizes artifacts based on their most recent update, distinguishing fresh, recently changed artifacts. It helps to compare their proportion against older and stable artifacts, which have not been touched in a long time.

\item \textbf{By Median Time:} As a hybrid of the previous two artifact orderings (Fig.~\ref{plot:plot_comparison}), sorting the artifacts by their median commit timestamp reveals the boundaries of the dataset both in terms of the creation of artifacts (bottom) and their gradual abandonment (top). A horizontal top boundary would indicate the presence of many recently updated artifacts, while a horizontal bottom boundary would show that many artifacts have been present in the repository since its inception.

\item \textbf{By Timestamps Similarity:} Groups artifacts with similar event distribution (comparing both absolute, relative or normalized timestamps), making it easier to detect artifacts with parallel evolution patterns, with many synchronous events.

\item \textbf{By Initial Metric Value:} Sorts based on the starting metric value, useful for identifying artifacts that began with specific characteristics.
\item \textbf{By Final Metric Value:} Orders artifacts by their most recent metric value, clustering artifacts with the same metric value in their latest version.
\item \textbf{By Metric Value Variation:} Highlights artifacts with the most substantial changes between the first and last recorded values, drawing attention to dynamic evolution.

\item \textbf{By Artifact Metadata:} Sorts artifacts by their path name, their filename extension (or both). The visualization reflects naming conventions introduced in the repository and can show refactorings involving changes in folder or filename, or the introduction of new programming languages in the code or migration across programming languages.
\end{asparaitem}

\subsection{Dot Color Encoding}

In \vizName, dot colors in the scatterplot can be used to encode additional artifact attributes:
\begin{asparaitem}
\item \textbf{Time (year):} The dots can be colored to represent the year in which the corresponding event occurred.
\item \textbf{Artifact type:} The color is mapped from the file extension, or other information that can be inferred from the artifact file name (e.g., documentation vs. code vs. test cases).
\item \textbf{Metric:} Uses a color gradient to represent the magnitude of a chosen metric, allowing for the identification of high and low values at a glance.
\item \textbf{Metric Variation:} Uses three colors to represent the positive, null or negative variation of a chosen metric, distinguishing additive changes vs. refactorings involving code removal and cleanup.
\item \textbf{Author:} Coloring based on the commit author may highlight different code ownership and collaboration patterns~\cite{10794911,girba2005developers}.
\end{asparaitem}

\subsection{Interactivity}

The visualization is meant to be enjoyed by interacting with it, by switching between an overview of the entire dataset and zooming into increasingly high levels of detail. To fit more data points into a limited set of pixels, it is possible to turn on the density plot feature, which uses the pixel opacity to visualize the density of data points under the visible pixel.

As the user changes the axis sorting criteria or time scaling configuration, it is possible to adjust the transition speed to smoothly shift between the two layouts. Tracking how the artifacts are reshuffled may reveal additional similarity patterns in the dataset. Users can control the transition duration, or turn the effect off to immediately switch between views.

The color mapping can also be changed interactively. Users can select which artifact feature (such as the file extension) or metric (such as the file size or the file size variation) is mapped to the color. The tool displays a histogram for each metric that 1) shows how many events fall under each class 2) provides a legend for the color mapping of the scatter plot. Users can also edit individual color mappings to highlight specific values in the scatter plot.

\begin{table*}[t]
    \caption{Gallery Dataset Size Overview}
    \setlength{\tabcolsep}{5pt}
    \label{tab:datasets}
    \centering
    \begin{tabular}{clcrrrrrrrr}
      \toprule
      ID & Dataset & Min. C/A & Artifacts & Commits & Events & Time Range & Years & First Timestamp & Last Timestamp \\
      \midrule
      D1 & Complete OpenAPI            & 0 & \numprint{256601} & \numprint{680115}  & \numprint{1374430} & \numprint{3868} days  & 10 & May 8, 2014, 06:22:41  & Dec 9, 2024, 15:40:37 \\
      D2 & Filtered OpenAPI          & 5 & \numprint{35441}  & \numprint{520754}  & \numprint{1045173} & \numprint{3865} days  & 10 & May 8, 2014, 06:22:41   & Dec 6, 2024, 21:41:47 \\
      \midrule
      D3 & GitHub Workflows   & 5 & \numprint{113816} & \numprint{1369110}                 & \numprint{2162788} & \numprint{1903} days  & 5  & Jul 26, 2019, 04:59:00                      & Oct 10, 2024, 17:17:22 \\
      \midrule
      D4 & Node.js               & 0 & \numprint{111728} & \numprint{42664}   & \numprint{612055}  & \numprint{5933} days  & 16 & Feb 16, 2009, 01:34:45  & May 16, 2025, 14:55:50 \\
      D5 & CPython            & 0 &  \numprint{24620}  & \numprint{111556}  & \numprint{335370}  & \numprint{12664} days & 34 & Sep 18, 1990, 12:47:40  & May 21, 2025, 13:10:57 \\
      D6 & VS Code             & 0 & \numprint{32823}  & \numprint{119580}  & \numprint{461367}  & \numprint{3471} days  & 9  & Nov 13, 2015, 15:18:17  & May 16, 2025, 01:36:04 \\
      D7 & Firefox & 0 & \numprint{864674}  & \numprint{787184}  & \numprint{6780112}  & \numprint{9911} days & 27 & Mar 28, 1998, 04:38:53  & May 16, 2025, 19:25:52 \\
      D8   & PostgreSQL         & 0 & \numprint{12098} & \numprint{60946}                 & \numprint{307195} & \numprint{10548} days  & 28  & Jul 9, 1996, 08:35:38                      & May 26, 2025, 13:30:01 \\
      \bottomrule
    \end{tabular}
\end{table*}

\subsection{Tool Architecture and Implementation}

\begin{figure}
    \centering
    \includegraphics[width=\linewidth]{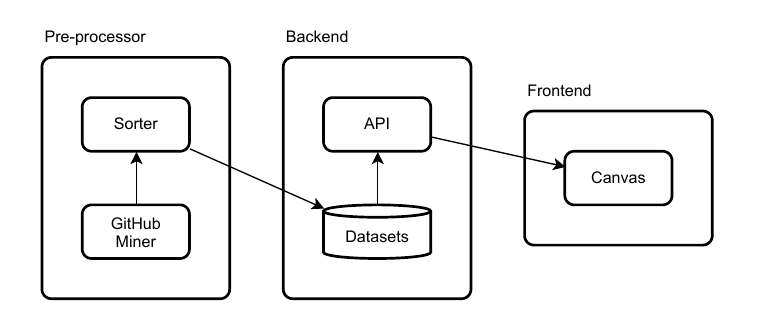}
    \vspace{-0.5cm}
    \caption{Architecture of \vizName}
    \label{fig:architecture}
\end{figure}

The tool is composed of a data crawler component (i.e., a GitHub crawler and git repository log miner), a pre-processor, the scatter plot canvas and its surrounding interactive controls (Fig.~\ref{fig:architecture}).

The pre-processor takes as input commit event logs, which should contain entries with a timestamp and an artifact identifier. It is possible to feed additional attributes or metrics extracted from the specific artifact version, such as for example its size, number of defects, the author of the commit affecting it. The role of the pre-processor is to prepare the dataset so that it can be efficiently rendered. This involves positioning each event in the dataset on the scatter plot coordinate system according to the various layout and sorting criteria discussed in the previous section. While some are very simple and -- even with a large dataset -- could be computed as the user switches between sorting criteria, others -- such as density clustering -- are very time consuming and should be pre-computed. To reduce storage space and bandwidth consumption, the results of the pre-processing are compressed.

The scatter plot canvas displays the pre-processed dataset after it is downloaded from the server and decompressed. It is also responsible for rendering the density plot, where -- due to the lower resolution of the display screen than what is required to paint the entire dataset, multiple events may overlap. This is reflected by their decreasing opacity level. The scatter plot also uses a spatial index to efficiently retrieve the artifact commit corresponding to the dot when the user mouse overlaps. The scatter plot canvas can also interpolate between two layouts and animate the transition with the give time.
A copy of the scatter plot image -- currently with the same display resolution as shown on the screen -- can be saved in PNG format.

To make it easy to reproduce a visualization, its configuration parameters are stored in a URL string that can be bookmarked or shared with other researchers.

\begin{figure}[t]
    \centering\footnotesize

    Node.JS \href{https://design.inf.usi.ch/evoscat/?dataset=node#time=absolute&artifact=mid&color=#000}{\includegraphics[width=.49\textwidth]{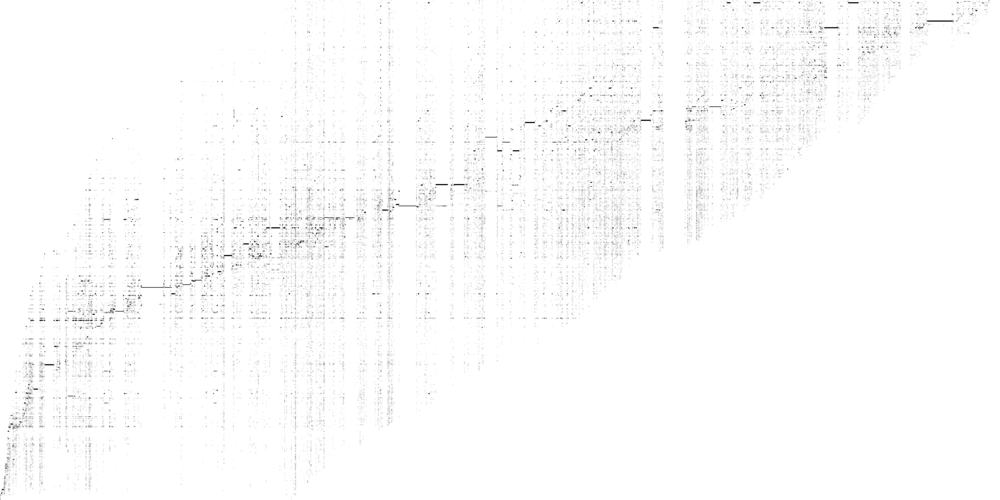}}

    CPython \href{https://design.inf.usi.ch/evoscat/?dataset=cpython#time=absolute&artifact=mid&color=#000}{\includegraphics[width=.49\textwidth]{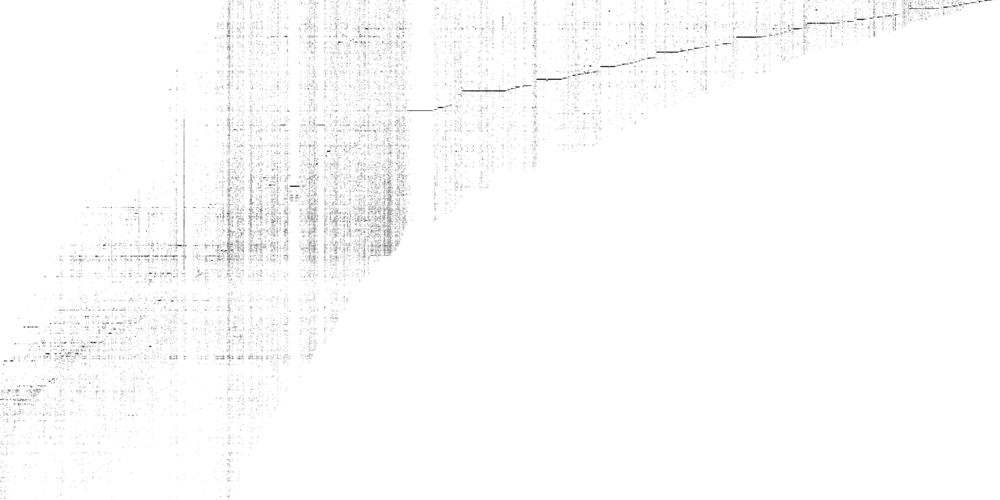}}

    VS Code \href{https://design.inf.usi.ch/evoscat/?dataset=vscode#time=absolute&artifact=mid&color=#000}{\includegraphics[width=.49\textwidth]{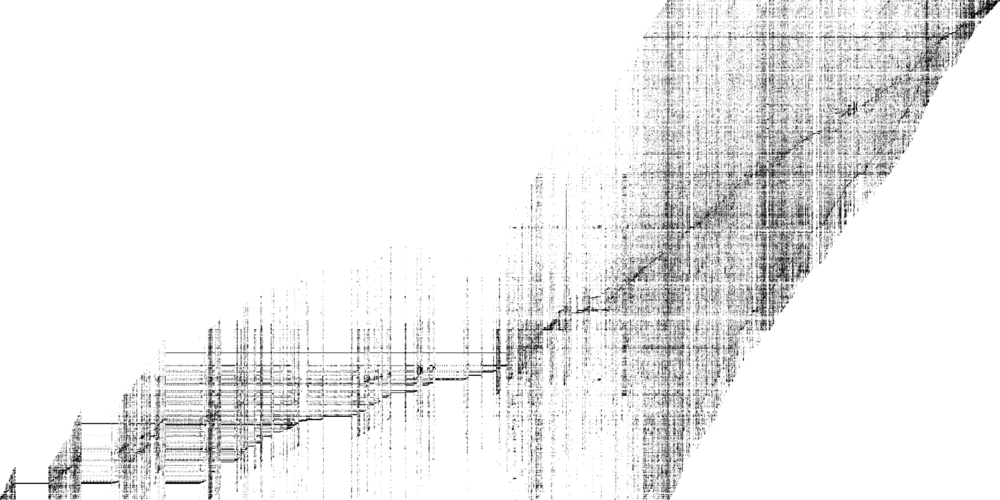}}

    Firefox \href{https://design.inf.usi.ch/evoscat/?dataset=firefox#time=absolute&artifact=mid&color=#000}{\includegraphics[width=.49\textwidth]{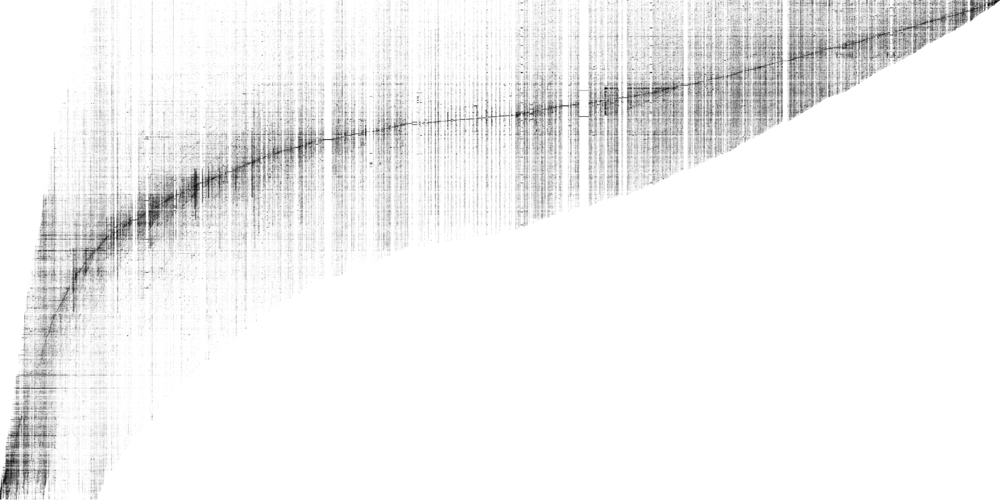}}

    \caption{Evolution Scatterplot Gallery: Artifacts sorted by median commit timestamp, Absolute time on the vertical axis. Black color intensity reflects density of commits and dataset size difference.}\vspace{-0.5cm}
    \label{plot:plot_comparison}
\end{figure}

\section{Gallery}\label{sec:gallery}

\subsection{Datasets}
We showcase the visualization in the case of eight datasets of commits ranging from nearly half a million events to more than six million events. An extended gallery\footnote{\url{https://design.inf.usi.ch/evoscat/gallery/}} containing more examples is available on \vizName website.

\subsubsection*{\textbf{D1: Web API Metrics Evolution Dataset}}

This dataset~\cite{apiace:2024:msr} traces the evolution of Web APIs through the changes detected in their corresponding specifications. For each specification commit, metrics related to the API size (in terms of functionality or data model) and the amount of endpoints covered by a certain security mechanism, in addition to other relevant metrics are computed. For this paper, we illustrate how \vizName is used to: (1) Give an overview on the usage of OpenAPI specification in public GitHub repositories; (2) illustrate the pace and frequency in which these files are being updated over time; In existing work~\cite{hurtado2025mining} we used the \vizName to (3) visualize and classify different Web API evolution trends based on their size and security coverage over time.

\subsubsection*{\textbf{D2: GitHub Workflows Evolution Dataset}}

The dataset is taken from the exploratory empirical analysis performed by ~\citeauthor{cardoen2024}~\cite{cardoen2024} on the histories of GitHub workflows. In particular, we have used a more recent version than the one used in the paper (Version 2024-10-25~\cite{cardoen2024dataset}). This updated dataset contains \numprint{2686974} workflow specifications extracted from \numprint{43342} different GitHub repositories. Note that we kept only all workflows with more than 5 commits, which made the dataset shrink to \numprint{2162788} ($-$\numprint{524186}) workflow specifications and \numprint{34150} ($-$\numprint{9192}) repositories.

\subsubsection*{\textbf{D3: Popular GitHub open source repositories}}
Unlike \textit{D1} and \textit{D2}, which comprise evolution histories of artifacts of the same type mined from many different software repositories, \textit{D3} contains the complete evolution history of all artifacts found within widely known open source repositories in GitHub. The goal is to showcase how \vizName can be used to represent the evolution dynamics of large and complex software repositories during their entire life time.

\subsection{Visualizations}

\subsubsection{Dataset overview (U1)}

Fig.~\ref{plot:plot_comparison} shows the overview of four different open source repositories, which have very different histories, for example, concerning the ratio of discarded artifacts (the highest in VS code) vs. the still active ones that had a long history (CPython) or a history spanning less than half of the project duration (Firefox). The four repositories also present a different tendency for the median commit of the artifacts. This is visible from the curve that emerges through the visualization that sorts the artifacts by their median commit time. A linear curve would indicate a regular rate of addition/removal of artifacts to the repository throughout the entire history. If the curve steepness increases (like in VS Code) the repository development pace picks up as more files are added to the repository.

Fig.~\ref{plot:ghwf_commit_absolute_time} shows various overviews angles from which the D2 dataset of events can be seen. It shows the strong visual effect of changing the criteria according to which the artifact are sorted in the x-axis, while keeping the time axis set to represent the absolute time of the commit in all plots.

In (a), artifacts are sorted by the timestamp of the last known event in the lifecycle of the artifacts. While it shows that many artifacts have activity within four years, the bottom right part of the plot is fading, which reflects that a very few number of the early GitHub Workflows are still being updated and maintained.

In (b), the artifacts are sorted by the timestamp of the first commit instead. The visualization reveals how early adopters initiated workflows, with more recent ones appearing in the upper part. This configuration helps highlight how quickly newer workflows are being adopted and whether older ones remain actively developed.

In (c), sorting artifacts by their median commit timestamp arranges them so that each column (vertical stack of dots) represents one artifact, with its central point of activity positioned along the horizontal axis. When the dense region of points aligns along a diagonal or tight band, it implies that for many artifacts, commits are clustered around their median timestamp  ---  meaning they are created and modified within a relatively short, consistent time frame.

In (d), the artifacts are sorted by the duration of their lifespan (age) --- from the shortest to the longest. One can see how longevity correlates with activity density. Short-lived workflows have tightly packed events, while longer-lived ones may show more sporadic updates. This view can be especially useful to differentiate between ephemeral automation setups and those that are maintained over extended periods.

\begin{figure}[t]
    \centering
    \href{https://design.inf.usi.ch/evoscat/?dataset=oas#time=normage&artifact=age&color=year}{\includegraphics[width=.99\linewidth]{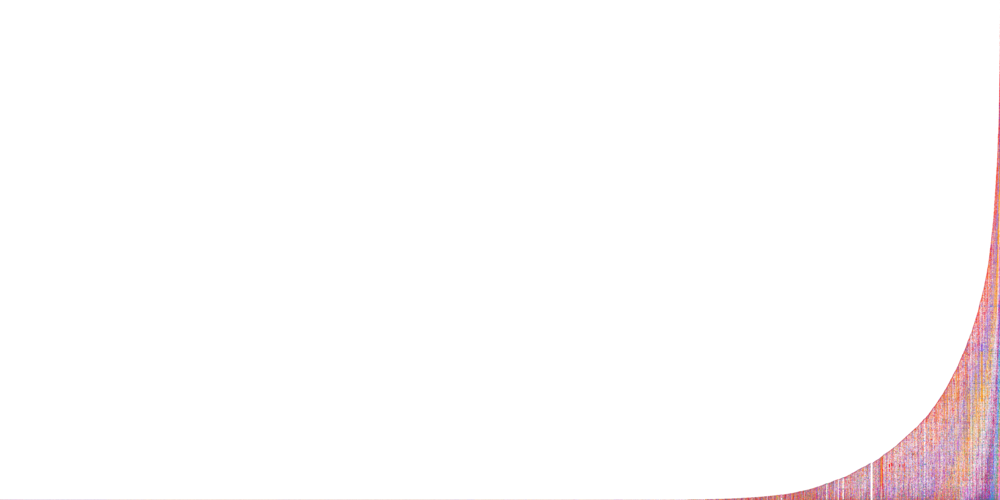}}

    \href{https://design.inf.usi.ch/evoscat/?dataset=oas5#time=normage&artifact=age&color=year}{\includegraphics[width=.99\linewidth]{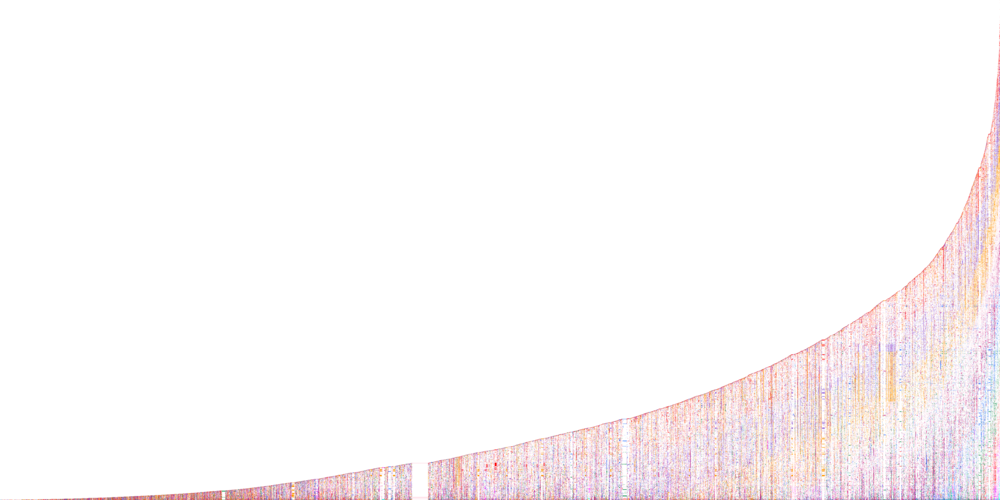}}

    \caption{Evolution Scatterplot of the complete OpenAPI Dataset (above) and filtered to include APIs with more than 5 commits (below) -- Artifacts sorted by their age, Time is relative to the first commit timestamp on the vertical axis, Colored by commit year}
    \label{plot:oas_complete_filter}
\end{figure}

\subsubsection{Artifact age distribution (U2)}

By sorting the artifacts by their age (the elapsed time between their last and first commit) and shifting the time axis to align the initial event of every artifact, we can compare the age distribution of artifacts across different datasets. For example, there is a clearly visible difference between the GitHub workflows dataset (Fig.~\ref{plot:ghwf_commit_absolute_time} (d)) and both the complete and filtered OpenAPI dataset (Fig.~\ref{plot:oas_complete_filter}).

Fig.~\ref{plot:oas_complete_filter} highlights that a very large proportion of the OpenAPI artifacts tend to be stable, as they have changed during a very short period of time after their creation or have been created but never updated afterwards. Using \vizName it is possible to filter those artifacts by setting the minimum number of events in their history.
Fig.~\ref{plot:oas_complete_filter} (bottom)
shows an example of the filtering outcome, where the plot only include the OpenAPI specification with more than 5 commits in their history, also shown in Fig.~\ref{plot:oas5-first-absolute} sorted by their first commit timestamp.

\begin{figure}[t]
    \centering
    \href{https://design.inf.usi.ch/evoscat/?dataset=vscode#time=absolute&artifact=index&color=year}{\includegraphics[width=.49\textwidth]{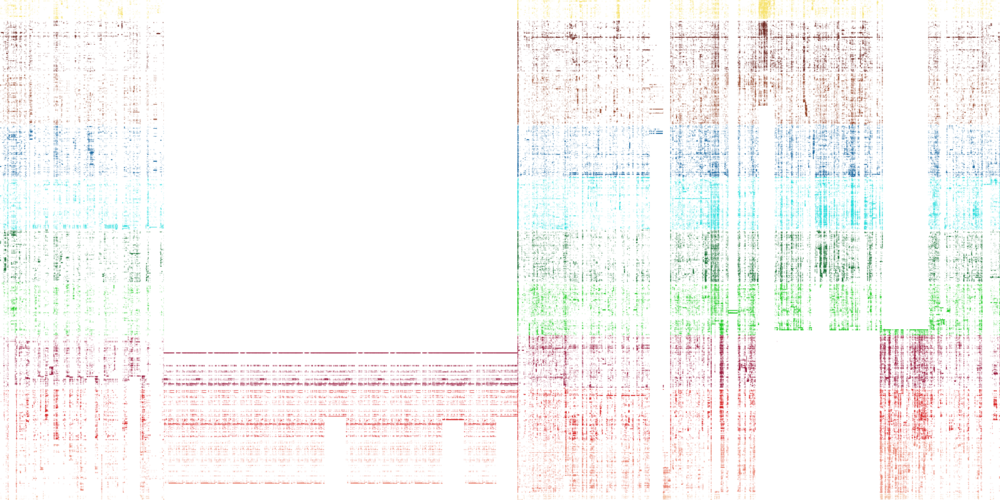}}

    \caption{Evolution Plot of the VSCode Dataset (Artifacts sorted by their pathname in the git repository, Time is absolute on the vertical axis, Colored by commit year) -- The discontinued, regularly spaced commits belong to the i18n/*.json files. The high intensity blot in the current year is the copilot integration. }
    \label{plot:vscode_id_absolute_time}
\end{figure}

\subsubsection{Development Pace and Effort Intensity (U3, U4)}

The regular structures found in the visualization of the VS Code dataset (Fig.~\ref{plot:vscode_id_absolute_time}) are due to changes applied systematically to the internationalization artifacts storing the translation of user interface elements in multiple languages. In the same Figure it is possible to spot a recently added set of files that have been the focus of a large number of changes in the most recent time period, which reflects the very intense effort invested in the integration of artificial intelligence features within the integrated development environment.

\begin{figure}[t]
    \centering
    \href{https://design.inf.usi.ch/evoscat/?dataset=cpython#time=absolute&artifact=short_extension&color=ext}{\includegraphics[width=.49\textwidth]{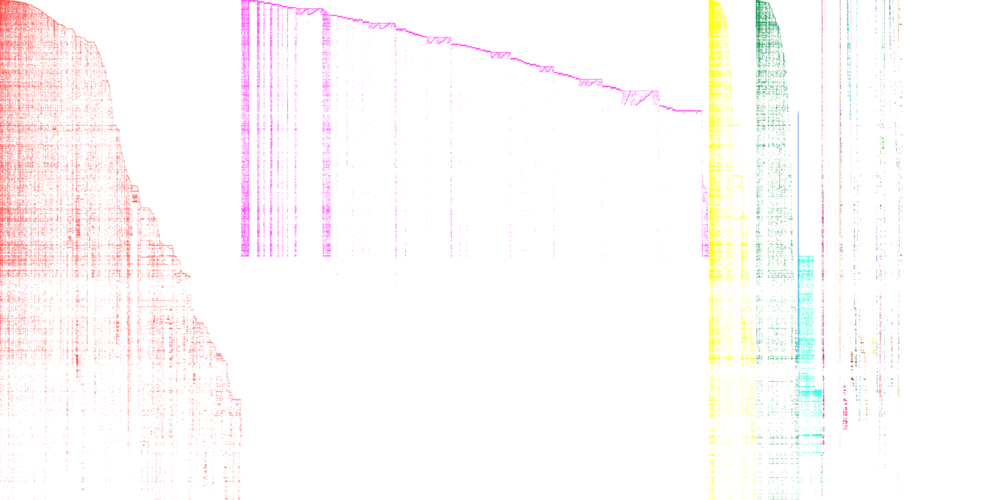}}

    \caption{Evolution Scatterplot of the CPython dataset. Artifacts sorted by 1) their extension 2) last commit timestamp, Time is absolute on the vertical axis, Colored by file extension -- The documentation originally written in latex \protect\legendbox{00FFFF} was migrated in one shot to use RST files \protect\legendbox{FF00FF} in the middle of August 2007. Later in September 2017 the NEWS file \protect\legendbox{0000FF} that had been continuously updated for 17 years was discontinued. From that moment onward, every news item was included in separate documentation files.}
    \label{plot:cpython_last_size}
\end{figure}

\begin{figure*}[!t]
    \centering
    \href{https://design.inf.usi.ch/evoscat/?dataset=wfgh#time=normtime&artifact=delta_methods_wave2_count&color=metric}{\includegraphics[width=.99\linewidth]{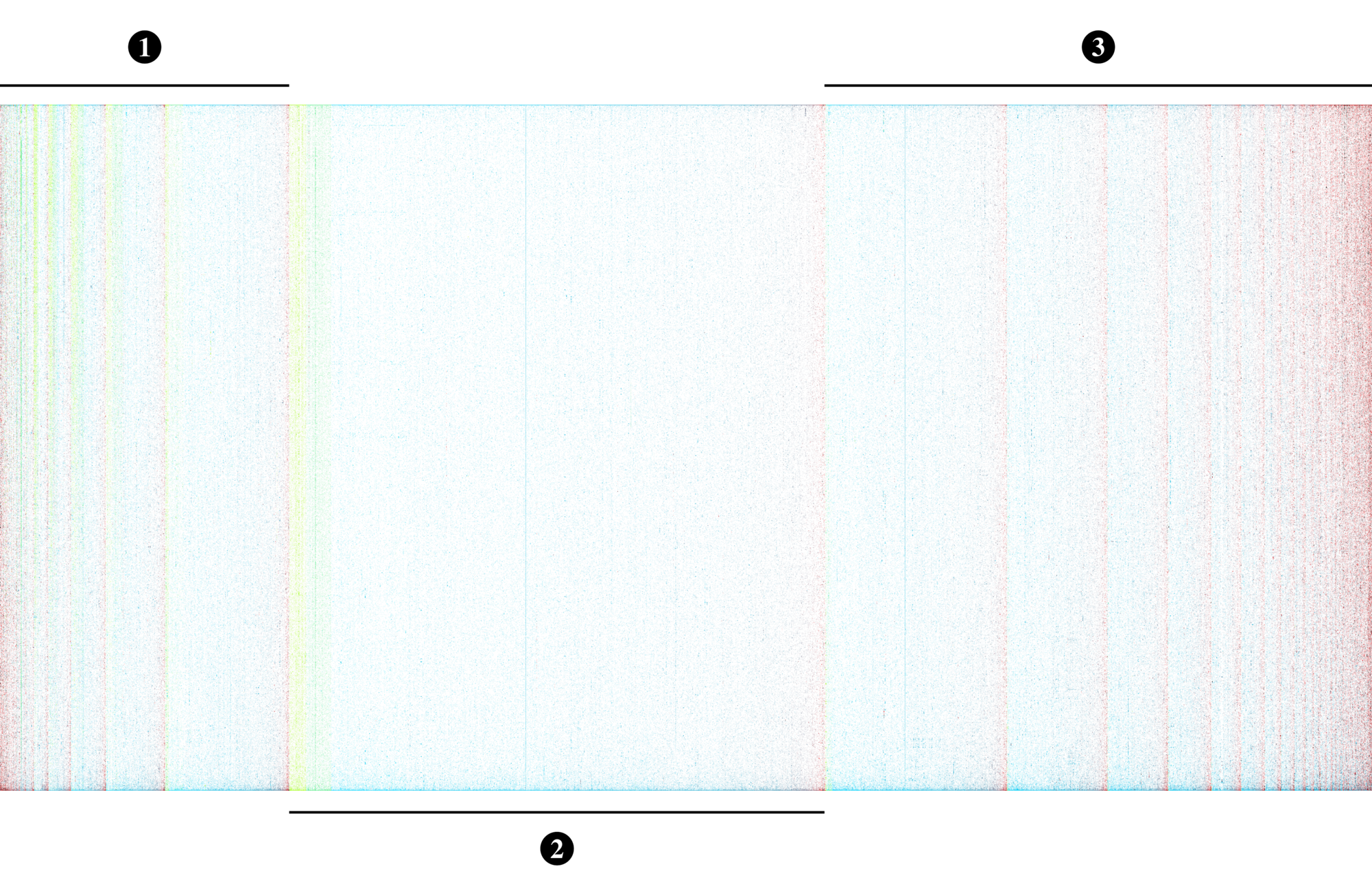}}
    \vspace{-0.5cm}
    \caption{Evolution Scatterplot of the Github Workflows dataset: Artifacts sorted by their 1) misconfiguration variation, 2) initial, and 3) final misconfiguration count, Time is normalized on the vertical axis, Colored by misconfiguration count: \protect\legendbox{00FF00} 0, \protect\legendbox{00FFFF} 2, \protect\legendbox{FF0000} 20, \protect\legendbox{000000} $>$ 20}
    \label{plot:ghwf_misconfiguration_delta_first_last}
\end{figure*}

\subsubsection{Stability vs. Metrics Variability (U5)}

Fig.~\ref{plot:ghwf_misconfiguration_delta_first_last} shows how the workflows' metric (in this case the number of security misconfigurations in the workflow) improves (i.e., less misconfigurations), stays the same, or worsens (i.e, more misconfigurations) over their histories. Each workflow evolution history is stretched along the entire plot height, the first commit being at the bottom, and the last on top. On the X axis we enumerate all the workflows in the dataset (sorted first by the variation of misconfigurations $\Delta M$ and then by the number of misconfigurations $M$). As we can see from the plot, we observe the formation of distinct vertical bands. The first group of bands \Circled{\textbf{1}} represents all the workflows for which the metric improves over time, \Circled{\textbf{2}} all those that stay the exact same, and finally \Circled{\textbf{3}} where the metric worsens over time.

The improvement and worsening of the metric in the workflows over time can be seen also by the color of each commit. In the case of the workflows on the left-side of the plot, we can see that the color goes from red (20 security misconfigurations) to cyan (2 security misconfigurations). Conversely, we can see that, for the workflows on the right-side of the plot, the colors go from cyan to red, and sometimes to black (more than 20 security misconfigurations).

The plot also shows that there is a relatively small group of commits containing 0 security misconfigurations (in green). In the case of the green commits in the first group of bands, it shows that all the misconfigurations were eradicated from the workflow in time. In the case of the green commits in the second group of bands, instead, it shows that there are some clean workflows that never had more than 0 misconfigurations throughout their whole history.

\subsubsection{Potential clone and fork detection (U6)}

Fig.~\ref{plot:oas5-first-absolute} shows how sorting by the initial commit timestamp helps to spot potential clone cases across different repositories due to the presence of repeated horizontal lines. The artifacts not only share the same initial commit timestamp (therefore, they are placed side by side in the visualization) but also share the same timestamp for many subsequent commits. In some of these cases, at some point in their history, the commits start having different timestamps, which could potentially indicate a fork.

\subsubsection{Interesting Patterns (U7)}

Sorting artifacts by filename extension presents the history of artifacts grouped by programming language and further separating code from documentation. As shown in Fig.~\ref{plot:cpython_last_size}, we can spot when the Python documentation system was migrated from latex to reStructuredText markup (RST) in one transition. In general, it is possible to spot transitions between artifacts (or types of artifacts) by tracking the presence of synchronous gaps in the plot, which indicate that some artifacts have been removed while at the same time other artifacts have appeared.

When using the absolute time axis, the presence of horizontal lines (Fig.~\ref{plot:vscode_id_absolute_time}) indicates commits affecting a large number of files, changing at the same time (e.g., upon a major release or refactoring). Vertical lines indicate artifacts which change with a high frequency (e.g., at the end of every working day).

When using the time axis centered around the median commit and the color mapping based on the metric variation, one can expect that the events representing artifact removal (or decrease in size) are found above the central line, while the events representing artifact creation (or changes that do not affect size) tend to be placed below.

\subsubsection{Artifact Freshness (U8)}

\begin{figure}[t]
    \centering

    \href{https://design.inf.usi.ch/evoscat/?dataset=postgres#time=absolute&artifact=double2_delta2&color=delta}{\includegraphics[width=.49\textwidth]{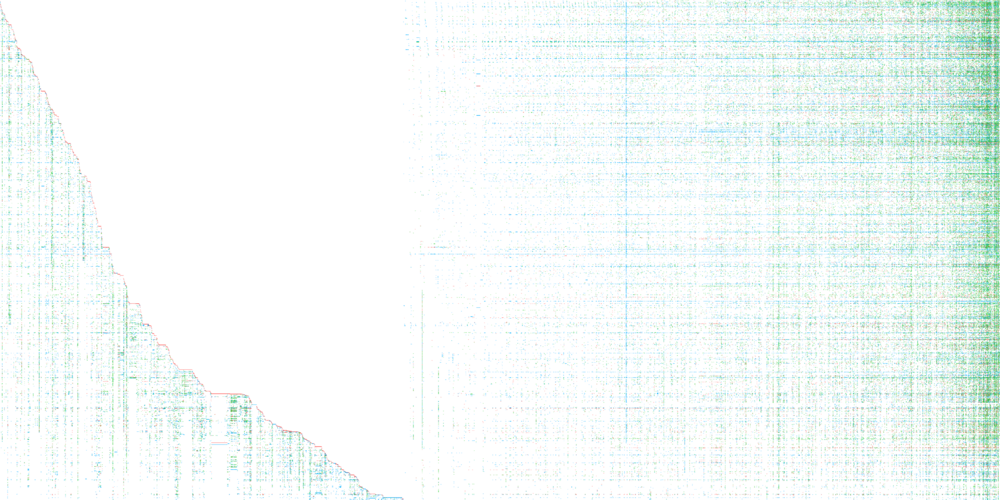}}

    \caption{Evolution Plot of the PostgreSQL Dataset (Artifacts sorted by 1) their size measured in the last commit 2) last commit timestamp, Time is absolute on the vertical axis, Colored by their size change \protect\legendbox{FF4A46} shrink or remove, \protect\legendbox{00AEFF} stable, \protect\legendbox{1fcb23} grow) -- Compared to Fig.~\ref{plot:vscode_last_size}, there are less discarded artifacts and a more ``punctuated'' evolution style with changes spanning across many artifacts.}
    \label{plot:evoscat-postgres-double2_delta2-absolute}
\end{figure}

 If we sort the VS code (Fig.~\ref{plot:vscode_last_size}) or PostgreSQL (Fig.~\ref{plot:evoscat-postgres-double2_delta2-absolute}) artifacts by their size as measured in the last commit and by the last commit timestamp. We can clearly distinguish two groups of artifacts. On the left, the ones which have been removed from the repository, while the currently active ones -- a much smaller group in case of VS Code -- are shown on the right.  This plot helps to visually estimate not only the proportion of the two kinds of artifacts but also the pace of artifact removal over the project lifetime.  This visualization may help to highlight major code rewriting efforts and support decision making on whether a historical analysis includes or excludes no longer present artifacts.

\subsubsection{Crawler Performance (U10)}

Empirical software engineering researchers build crawlers to gather artifacts from open source repositories. Given the rate limit imposed by GitHub and other sources, the crawlers need to balance two different goals: 1) discovery of new relevant artifacts, 2) tracking the artifact evolution history. By comparing Figs.~\ref{plot:oas5-first-absolute} and~\ref{plot:ghwf_commit_absolute_time} (b), we can see that the GitHub workflow crawler discovered new artifacts at a constant rate throughout the years, while in case of the OpenAPI dataset, the priority switched in mid 2023 from new artifact discovery to tracking the history of existing artifacts, as reflected in top-right part of Fig.~\ref{plot:oas5-first-absolute}.

\section{Limitations}\label{sec:discuss}

The main limitations of the visualization concern the source of events, which are currently taken from the main branch of a repository. While one could easily try to represent branches as additional artifacts side by side with the ones taken from the main branch, git repositories do garbage collect commits belonging to deleted branches, which are either discarded (and lost from the perspective of empirical software evolution researchers) or eventually merged with the main branch (and therefore visible in the current visualization).

Since each artifact is identified by its path name, we do not explicitly track renaming of artifacts, which is shown as the removal of the previously named artifact followed by the addition of a new artifact with the new name.

We also assume that the timestamp provided by the git log is reliable, which may not always be the case~\cite{flint2022pitfalls}. The visualization has also been used to spot the presence of out of range commit timestamps (too old, or in the future) and may thus help to motivate the need for data cleanup or inspect the results against the input of the data cleanup phase.

There are some fundamental limits on the raw size of the datasets that can be visualized with the current tool implementation. These concern the maximum amount of data that can be fetched as JSON strings by Web browsers, as well as the increasingly long amount of time necessary to pre-process the data. The limited resolution of the scatter plot canvas also imposes a limit on the maximum number of artifacts that can be shown on the same plot.

\section{Related Work}\label{sec:relw}

\subsection{Visualizing large datasets}

Different solutions are presented to address scalable scatterplot visualizations as \citeauthor{tao2020kyrix} presented Kyrix-S\cite{tao2020kyrix}, a tool that relies on multinode database storage and the use of multinode spatial indexes to achieve interactive browsing of large scatterplot visualizations, which can scale to represent datasets with one billion Reddit comments. Other projects such as \citeauthor{hilasaca2023grid} \cite{hilasaca2023grid}, presented a strategy to remove overlaps from scatterplot layouts with dimensionality reduction, to enhance readability. However, they consider that the elimination of overlaps affects the distribution represented in the plot, an issue due to overlap removal techniques, such as DGrid and ReArrange, and also can lead to information loss in the visualization. The work by \citeauthor{liu2021visual} \cite{liu2021visual} presented a visual analytics system to support evolution analysis using topic modeling. They employ the Latent Dirichlet Allocation (LDA) technique to analyze source code from open-source projects on GitHub as JavaScript libraries, D3.js, and Vue.js.
After characterizing the software features and classifying the source files, they quantify differences between versions to reveal evolution patterns and development pace.

In contrast to these projects, which are focused on large-scale historical data but are limited to scaling within specific experimental contexts,
we introduce \vizName, a tool that displays large-scale visualizations that dynamically adapt to diverse datasets with rich historical data.

\subsection{Software evolution visualization}

\citeauthor{Pfahler2020Cities}~\cite{Pfahler2020Cities} extended the city metaphor, in which buildings represent classes and districts represent packages by \citeauthor{wet2007vis}~\cite{wet2007vis} to visualize evolving software systems.  To do so, they introduced new ways to traverse time through a ``History-Resistant Layout'', in which a fixed position in the 3D space is given to each element of the city. By doing so, they managed to remove the unpredictable jumps of buildings and districts during the progression of its history, making for a more cohesive and easy-to-follow animation. The evolution scatterplot needs to deal with similar layout issues to position artifacts along the X axis, a mono-dimensional space. As the time dimension is explicitly represented, we use animations for the transition between different layouts.

\citeauthor{Sandoval2019Performance} introduced the Performance Evolution Matrix~\cite{Sandoval2019Performance}, an interactive matrix visualization where software components are arranged as rows and software versions as columns. Each cell encodes run time metrics, enabling developers to trace performance regression or improvement across versions. Listing components along the x-axis provides a comparative view, helping to spot the component that changed performance the most in absolute or relative terms. This comparative view can be also provided by \vizName at scale: Fig.~\ref{plot:ghwf_misconfiguration_delta_first_last} provides a visual comparison of how many CI/CD workflows reduce, increase or have a stable number of misconfigurations detected throughout their whole history.

Similar comparisons can be made with CCEvovis~\cite{Honda2019}, a tool that uses stacked barcharts to track clone changes across versions. Each bar represents a specific version of the software. The height of the bar corresponds to the total number of clone sets detected in that version. The bar is divided into colored segments that indicate the number of clone sets belonging to each evolution category within that version: added clone sets (red segment), changes clone sets (green segment), and deleted clone sets (dark blue). Looking at the bars next to each other allows comparing the clone sets across software versions. When using \vizName to visualize artifact histories mined from different repositories (Fig.~\ref{plot:oas5-first-absolute}) the presence of potential clones with identical evolution histories can be visually spotted through repeated horizontal lines.

\begin{figure}[t]
    \centering
    \href{https://design.inf.usi.ch/evoscat/?dataset=postgres#time=absolute&artifact=mid&color=delta}{\includegraphics[width=.49\textwidth]{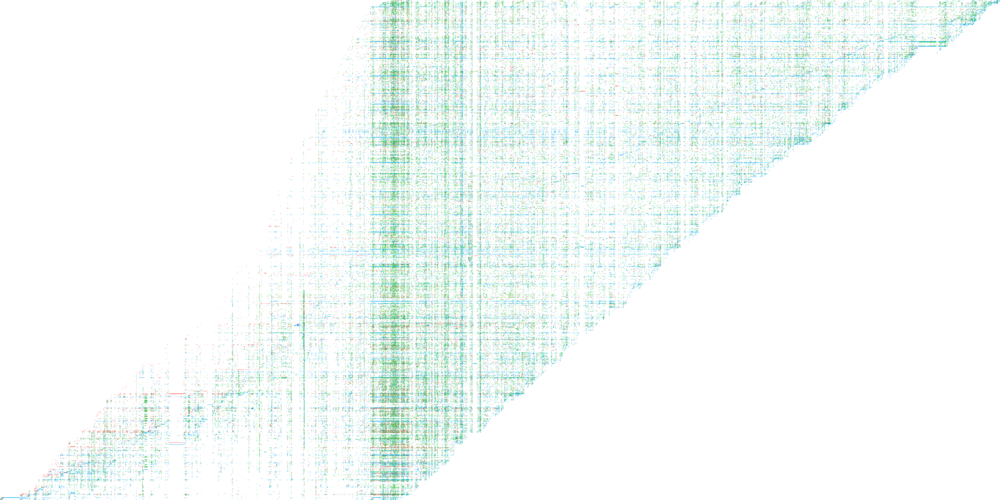}}

    \caption{Evolution Plot of the PostgreSQL Dataset. Artifacts sorted by their median commit time, Time is absolute on the vertical axis, Colored by their size change \protect\legendbox{FF4A46} shrink or remove, \protect\legendbox{00AEFF} stable, \protect\legendbox{1fcb23} grow -- Through the entire project there has been a long lasting core of frequently changing artifacts to which new ones are added and removed following a regular pace.}
    \label{plot:evoscat-postgres-mid-absolute-delta}
\end{figure}

Pioneering work on the visualization of software evolution includes RelVis~\cite{pinzger2005visualizing} by \citeauthor{pinzger2005visualizing} and evolution spectrographs \cite{wu2004evolution} by  \citeauthor{wu2004evolution}, providing graphical representations of source code and release history data. RelVis represents software modules, in this case, the nodes represent source code entities, edges the relationships between them, and the evolution of the metrics by highlighting the resulting polygons. Evolution spectrographs are used to visualize and track architectural changes, specifically modifications at the file level, in response to new entry requirements. They introduced evolution spectrographs to study changes in specific open source systems: OpenSSH, PostgreSQL (Figs.~\ref{plot:evoscat-postgres-double2_delta2-absolute} and~\ref{plot:evoscat-postgres-mid-absolute-delta}), and Linux. The goal was to gain a better understanding how software evolves by looking for evidence of specific changes. In contrast with our study based on the visualization of software artifact histories, our work not only represents the temporal evolution of the artifacts but also gives the possibility to explore their similarity and long-term evolutionary trends. Particularly regarding the PostgreSQL dataset, we are glad to report that the original finding from more than 2 decades ago of a ``punctuated evolution'' with a linear rate of growth~\cite{wu2004evolution} appears to have been maintained during most of the entire project lifetime (Fig.~\ref{plot:evoscat-postgres-mid-absolute-delta}).

\section{Conclusion}\label{sec:theend}

In this paper we introduce \vizName. A tool that allows producing interactive and scalable visualizations attempting to help researcher compare historical datasets, tracking the evolution of a large number of artifacts over long periods of time. \vizName's visualization is designed to contain millions of events, such as commits that create, change or remove artifacts from one or multiple code repositories. Users can not only zoom to focus on specific artifacts, but also can rearrange how they are sorted along the artifact axis and configure how colors are mapped based on different metrics or features associated with the artifacts. We included a gallery of visualizations illustrating the history of multiple popular open source repositories as well as comparing specific artifacts (such as GitHub workflows and OpenAPI specifications) mined from thousands of repositories.

In the future we plan to feature a side-by-side comparative visualization of different repositories as well as to make it easier to merge datasets, e.g., to study all repositories managed by the same organization.
We have also started to experiment with replacing the time axis with a different one, using a different metric to position the events along the history of artifacts, revealing an interesting relationship between the size of an API and its security coverage~\cite{hurtado2025mining}. We are also working on the integration of the evolution scatter plot with existing detailed visualizations of specific artifact evolution histories~\cite{apiace:2023:vissoft}.

\section{Supplementary Material}

A demo video of \vizName is available at \url{https://youtu.be/z_SLstI1mx8}. The source code and datasets are published at \url{https://zenodo.org/records/15525004}.

\clearpage
\begin{small}
\bibliography{bib}
\bibliographystyle{plainnat}
\end{small}
\end{document}